# Improvement of infrared single-photon detectors absorptance by integrated plasmonic structures

M Csete[1,2], Á Sipos[2], A Szalai[2], F. Najafi[1], G. Szabó[2], K. K. Berggren[1]

[1] Massachusetts Institute of Technology, Research Laboratory of Electronics, 77 Massachusetts Avenue, Cambridge, Massachusetts, USA 02139

[2] University of Szeged, Department of Optics and Quantum Electronics, H-6720, Szeged, Dóm tér 9, Hungary

**Abstract:** The absorptance of p-polarized light in superconducting-nanowire single-photon detectors (SNSPDs) was improved by integrating (1) ~quarter-wavelength nano-optical cavity closed by a gold reflector (OC-SNSPD), (2) nano-cavity-array closed by vertical and horizontal gold segments (NCAI-SNSPD), and (3) nano-cavity-deflector-array consisting of longer vertical gold segments (NCDAI-SNSPD) into short- ($p$-) and long- ($3p$-) periodic niobium-nitride (NbN) stripe-patterns. In OC-SNSPDs the highest absorptance is observable at perpendicular incidence onto NbN stripes in P-orientation due to E-field concentration at the bottom of nano-cavities. In short-periodic NCAI-SNSPDs off-axis illumination results in almost polar-angle-independent perfect absorptance due to collective resonances on plasmonic MIM nano-cavity-arrays in S-orientation. In long-periodic NCAI-SNSPDs the surface wave-excitation phenomena promote EM-field transportation to the NbN stripes in S-orientation and results in local absorptance maxima. In NCDAI-SNSPDs with proper periodicity large absorptance maxima appear due to synchronous E-field enhancement via deflected SPPs below NbN stripes in S-orientation, which make possible fill-factor-related loss compensation.

The near-field enhancement accompanying the propagating and localized plasmons is widely applied in different areas of nano-photonics including out-coupling from LEDS [1], in-coupling into photo-detectors [2] and bio-sensors [3], emission enhancement and laser beam-shaping [4, 5, 6], and high harmonic generation [7]. The complex plasmonic structures are the most promising, as the highest degree of freedom is attainable in spectral engineering due to the interaction of supported localized and propagating surface modes [7, 8].

An important application area of plasmonic structures is the enhancement of infrared detectors' efficiency, e.g. absorptance of superconducting nanowire single-photon detectors (SNSPDs) [9]. Novel approaches based on plasmonic structures' integration into SNSPDs open novel avenues also in secure communication and in quantum-key computing [10].

The most simple noble metal structure integrated into SNSPDs was a thin gold reflector aligned onto the top of quarter wavelength dielectric cavity, which made possible to reach ~50 % absorptance [11, 12]. Even though the propagating plasmonic modes are close to the light line in the IR spectral region, our previous studies have proven that it is possible to concentrate the EM-field around the nanometric niobium-nitride (NbN) stripes, which are the absorbing elements in SNSPDs, via resonant nano-plasmonic modes [13-15]. The advantage of appropriately designed complex integrated nano-antennas is the squeezing of the IR light around the tiny absorbing NbN segments in SNSPDs [13, 14]. The expelling of the **E**-field from noble metal and the coupling to localized plasmonic modes enabled to attain ~95.7 % absorptance in short- (200 nm) periodic, and ~46.5% absorptance in long- (600 nm) periodic nano-cavity-array integrated SNSPD designs [14, 15].

Besides the device geometry development, the effect of illumination direction was also investigated, and the possibility of absorptance enhancement via total internal reflection (TIR), Brewster and plasmonic phenomena was demonstrated [13-16]. It was shown that in noble metal integrated SNSPDs those orientations are the most efficient, where the **E**-field oscillation direction is perpendicular to the integrated plasmonic elements [13-15].

In our present study the optimal illumination directions were determined for three different SNSPD device designs consisting of (1) ~quarter-wavelength nano-optical cavity closed by a gold reflector (OC-SNSPD, Fig. 1a), (2) nano-cavity-array closed by vertical and horizontal gold segments (NCAI-SNSPD, Fig. 1b), and (3) nano-cavity-deflector-array consisting of longer vertical gold segments (NCDAI-SNSPD, Fig. 1c). Two periodicity intervals were investigated, first the $p$~200 nm periodicity conventional in SNSPD devices was studied, where the detection efficiency is optimized via absorptance maximization; and then the larger $3p$~600 nm periodicity capable of realizing simultaneous electrical optimization was inspected, similarly to ref. [15]. The purpose of our present work was to determine the optimal geometrical parameters and the optimal orientations for these device designs, and the idea we applied is **E**-field concentration via localized and propagating plasmonic modes.

Finite element method was applied, and similar three-dimensional models were prepared following the numerical approach described in our previous paper [11]. The optimal illumination directions were determined and wavelength dependent computations were also performed at specific orientations of each designs, where NbN absorptance maxima are predicted based on angle dependent studies. The geometry of the NbN stripes is in accordance with conventional SNSPDs, but the geometry of the NbN pattern was also varied for each three designs, namely three different periodicities were inspected in the short $p$- and long $3p$-periodicity regions in order to ensure the highest absorption cross-section, and reduced kinetic inductance, simultaneously.

In OC-SNSPD device designs the NbN pattern is embedded into ~quarter-wavelength, nano-optical cavity filled with dielectric and closed by a gold reflector (Fig. 1a).

The OC-SNSPD is similar to the optical system described in [11], but in present work three different periodicities are investigated in both *p* and 3*p* periodicity regions.

In NCAI-SNSPD device designs the NbN stripes are aligned at the bottom of ~quarter-plasmon-wavelength nano-cavities, which are closed by vertical and horizontal gold segments, and compose plasmonic MIM cavity-array with *p* and 3*p* periodicity (Fig. 1b).

The entirely novel NCDAI-SNSPD device design is based on nano-cavity-deflector-array consisting of longer vertical gold segments. In order to minimize the spurious gold absorptance, as well as to maintain the commensurability of the periodicity with the wavelength of the surface waves, deflector segments with 3*p* periodicity were integrated into both *p* and 3*p* periodic NbN patterns (Fig. 1c).

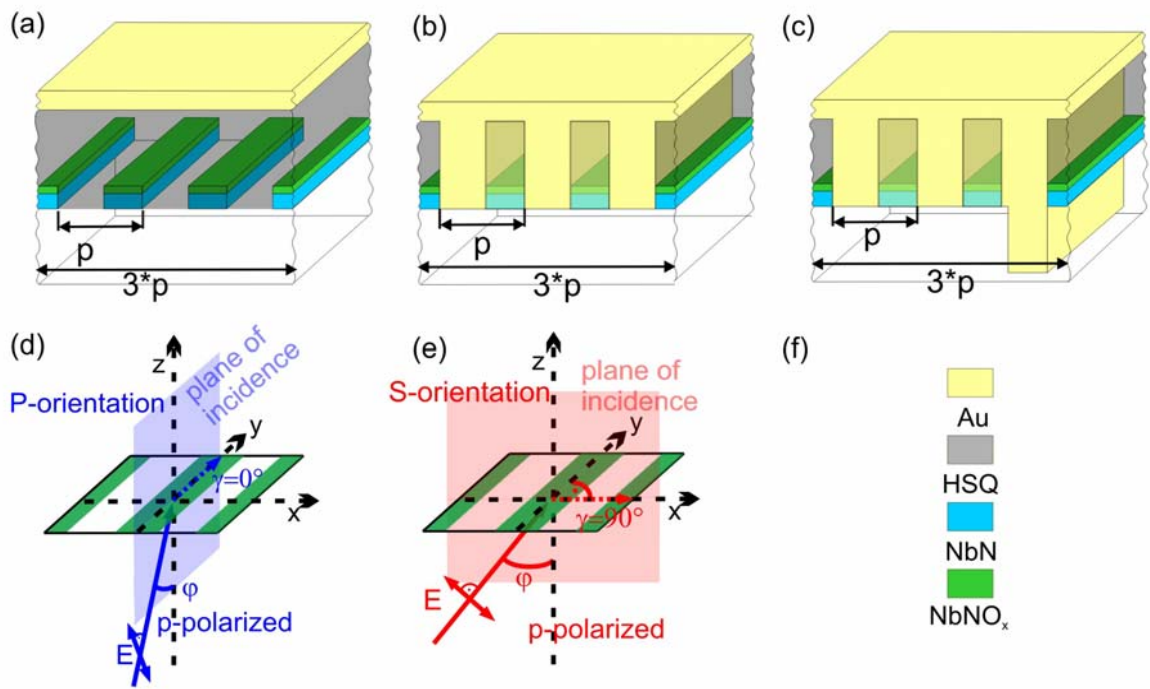

Figure 1. Schematic drawings of device designs with periodicities *p*= 200 nm, 220 nm and 237 nm / 3*p* = 600 nm, 660 nm and 710 nm: (a) OC-SNSPD, (b) NCAI-SNSPD, (c) NCDAI-SNSPD (in 3*p* periodic patterns the segments indicated by pale-yellow are also filled with gold). The configuration applied to realize p-polarized illumination in (e) P-orientation, (f) S-orientation, and the identification of materials in the integrated SNSPD devices.

**Results**

The inspection of the dual-angle-dependent absorptance of NbN patterns revealed that larger absorptance is achievable due to p-polarized illumination of NbN patterns in P-orientation ($\gamma=0°$) in all OC-SNSPDs, while higher absorptance is achievable in S-orientation ($\gamma=90°$) in all NCAI-SNSPD and NCDAI-SNSPDs. The absorptances in these orientations were compared for different periodicities of each specific design (Fig. 2, 3a and 4, 5a). The inspection of the spectral effects at the maxima on the polar angle dependent NbN absorptance of each designs indicated the particularity of a specific $3p$ periodicity (Fig. 3b and 5b). The investigation of the normalized **E**-field distribution at the global extrema uncovered the role of localized and propagating plasmonic modes in absorptance maximization (Fig. 6 and 7). The maximal absorptance values correlate with the resistive heating in the lossy NbN stripes, as it was shown in [11, 15, 17].

**NbN absorptance in short-periodic designs**

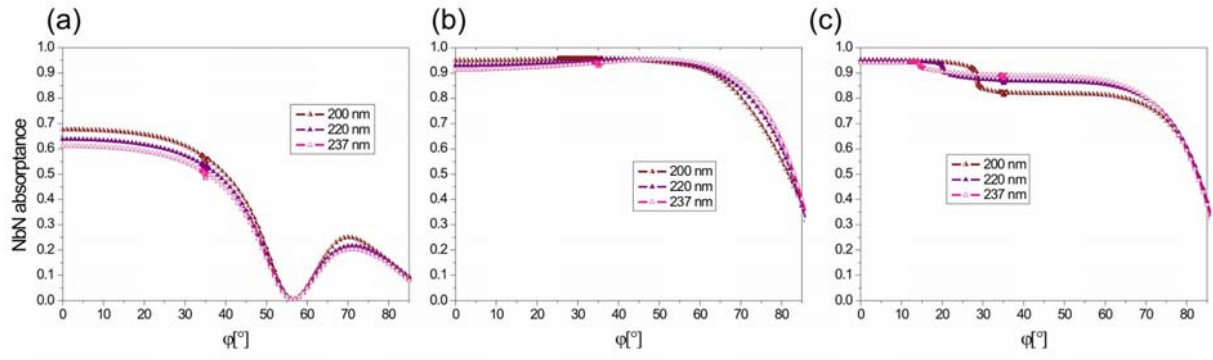

Figure 2. NbN absorptance in three different short-periodic SNSPD designs: (a) OC-SNSPD in P-orientation, (b) NCAI-SNSPD in S-orientation, and (c) NCDAI-SNSPD in S-orientation.

In all three short-periodic OC-SNSPDs the maximal absorptances (67.6%, 63.8%, 61.1%) are observable at perpendicular incidence (Fig. 2a). Local perturbations appear at the orientations corresponding to the condition of TIR and surface plasmon polariton (SPP) excitation phenomena. Local maxima are reached at the same ~71° tilting, which originate from NbN-related ATIR phenomenon. All global and local maxima decrease with the fill-factor in $p$ periodic OC-SNSPDs.

In short-periodic NCAI-SNSPDs almost tilting independent perfect absorptance is observable through large polar angles (Fig. 2b).

The TIR and SPP phenomena manifest themselves in local perturbations also in these designs at 34.5° and 35.1° polar angles. Interestingly, the 95.7% global absorptance maximum appears at 34.55° in 200-nm-pitch design, i.e. neighboring the orientation corresponding to TIR phenomenon, while it is shifted to larger 44° (95.4%) and 50° (95.3%) polar angles in the 220-nm and 237-nm-pitch designs. The global maxima decrease with the fill-factor also in *p* periodic NCAI-SNSPDs.

In short-periodic NCDAI-SNSPDs the maximal absorptances (95.1%, 94.9%, 94.1%) are reached at perpendicular incidence, then the absorptance monotonously decreases with the tilting (Fig. 2c). Steep absorptance changes are observable at polar angles, which slightly shift backward, i.e. to smaller polar angles, with the increase of the periodicity. These orientations correspond to the global maxima observable in corresponding long-periodic NCDAI-SNSPD designs (See next section). In 237-nm-pitch design the 94.2% absorptance at the particular orientation slightly overrides the absorptance observable at perpendicular incidence, but the global maxima decrease with the periodicity also in *p* periodic NCDAI-SNSPDs.

The normalized NbN absorptance monotonously increases in short-periodic devices by increasing the periodicity (Fig. 3), with the smallest slope and values in OC-SNSPDs, i.e. these devices are the less promising in absorptance maximization. Commensurate NbN absorptances are reached in NCAI-SNSPDs and NCDAI-SNSPD with the same periodicities, but the slope of the normalized absorptance curve is larger in presence of deflectors suggesting that application of larger periodic patterns is more beneficial due to decrease of the gold-to-NbN ratio. No peculiar pitch is identifiable in the interval p-periodicities, according to their sub-wavelength nature.

The absorptance in *p* periodic designs monotonously increases with the wavelength at the global maxima, which appear at perpendicular incidence in OC-SNSPDs and at polar angles increasing with the periodicity (34.55°, 45°, 50°) in NCAI-SNSPDs. The wavelength dependent absorptance in *p* periodic NCDAI-SNSPDs indicates similar steep slope, as it was observed in polar angle dependent studies, but here the inflection point forward shifts with the periodicity, according to the inverse proportionality of the wave vector to the wavelength. Sudden slope modification occurs in the closest proximity of 1550 nm in 220-nm-pitch design.

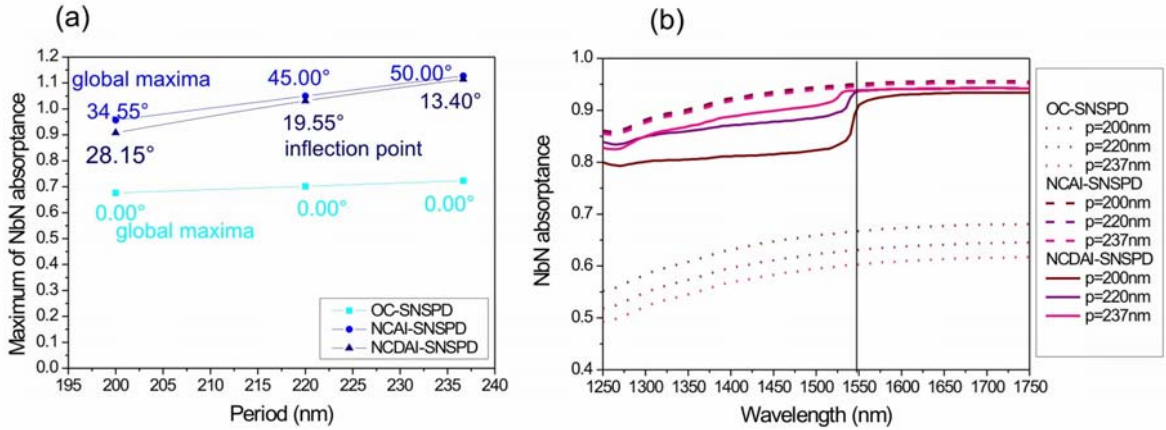

Figure 3 (a) The normalized absorptance in *p*-periodic SNSPD designs (b) wavelength dependent absorptance at the orientations indicated in (a) in different SNSPD designs.

**NbN absorptance in long-periodic designs**

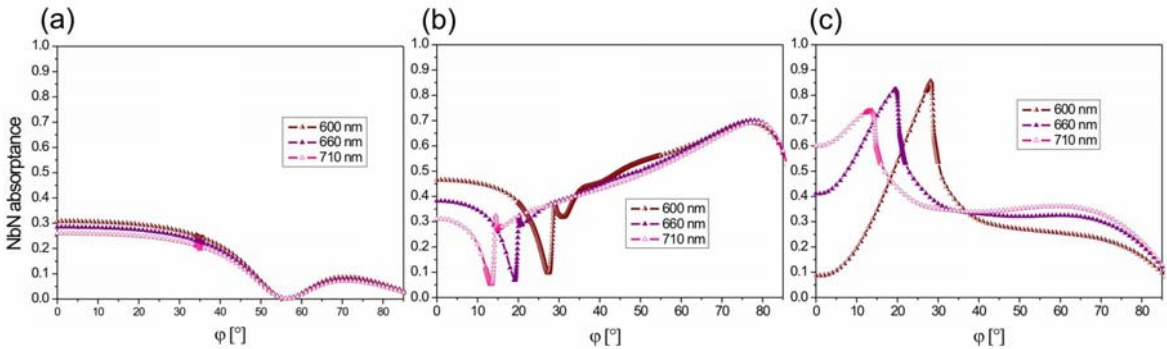

Figure 4. NbN absorptance in three different long-periodic SNSPD designs: (a) OC-SNSPD in P-orientation, (b) NCAI-SNSPD in S-orientation, and (c) NCDAI-SNSPD in S-orientation.

In all three long-periodic OC-SNSPDs the maximal absorptance (30.7%, 28.5%, 25.9%) is reached at perpendicular incidence (Fig. 4a). Local perturbations appear originating from TIR and SPP excitation phenomena at the same tilting, as observed in short-periodic designs. The NbN-related ATIR results in appearance of small local maxima at ~71° polar angle, analogously with the short-periodic OC-SNSPDs. Three-times reduced fill-factor causes considerably smaller global maxima, than in short-periodic OC-SNSPDs, which decrease with the fill-factor.

In long-periodic NCAI-SNSPDs the absorptance is significantly enhanced (46.5%, 38.2%, 31.1%) at perpendicular incidence in comparison to the OC-SNSPDs with same pitch (Fig. 4b-to-a). The tilting causes sudden changes in the absorptance, which manifests itself in global minima followed by local maxima in 3*p* periodic NCAI-SNSPDs. Interestingly, the course of absorptance is very similar in all three systems, but the position of the global minima and local maxima shifts backward, i.e. to smaller polar angles, with increasing periodicity.

The local maxima (37.2%, 32.3%, 32.3%) appear at 29°, 20.2° and 14.5° polar angles, indicating that their appearance is governed by grating-coupling phenomenon. The position of the global absorptance maxima (69.1%, 70.1%, 69.3%) is not influenced by the periodicity, on the contrary, these maxima appear at the same ~78° polar angle corresponding to NbN-related ATIR phenomenon. The global maximum is the highest in 660-nm-pitch design, i.e. the maximal absorptance is a non-monotonous function of the fill-factor in NCAI-SNSPDs.

In long-periodic NCDAI-SNSPDs the absorptance is strongly reduced at perpendicular incidence, then suddenly increases with the tilting through large global maxima (85.5%, 82.3%, 73.8%), which approximate the absorptances in corresponding short-periodic designs (Fig. 4c-to-2c). These global maxima backward shift with the increase of the periodicity, namely they appear at 28.15°, 19.55°, 13.4°, i.e. coincide with the inflection points observable on the absorptance curves of corresponding short-periodic NCDAI-SNSPDs. The absorptance decreases with the fill-factor also in long-periodic NCDAI-SNSPDs.

The normalized NbN absorptance is commensurate in OC-SNSPDS and NCAI-SNSPDs, while it is ~3 times larger in NCDAI-SNSPDs (Fig. 5a). Interestingly, the normalized absorptance indicates a local maximum at 660-nm-pitch already in OC-SNSPD. In NCAI-SNSPD designs this periodicity is not advantageous at the tilting corresponding to the local maximum, on the contrary, the normalized NbN absorptance indicates a local minimum at 660-nm-pitch. In $3p$ NCDAI-SNSPDs, although the maximal absorptance decreases with periodicity, the normalized NbN absorptance indicates a maximum at 660 nm, revealing to the particularity of 660-nm-pitch NCDAI-SNSPDs.

The wavelength dependent absorptances in $3p$-periodic designs indicate characteristics, which differs from the polar angle dependent absorptance according to the inverse proportionality of the wave vector to the wavelength. In OC-SNSPD the absorptance monotonously increases with the wavelength, there are no local changes at 1550 nm. In NCAI-SNSPDs the local absorptance maximum appears exactly at 1550 nm in 660-nm-pitch design.

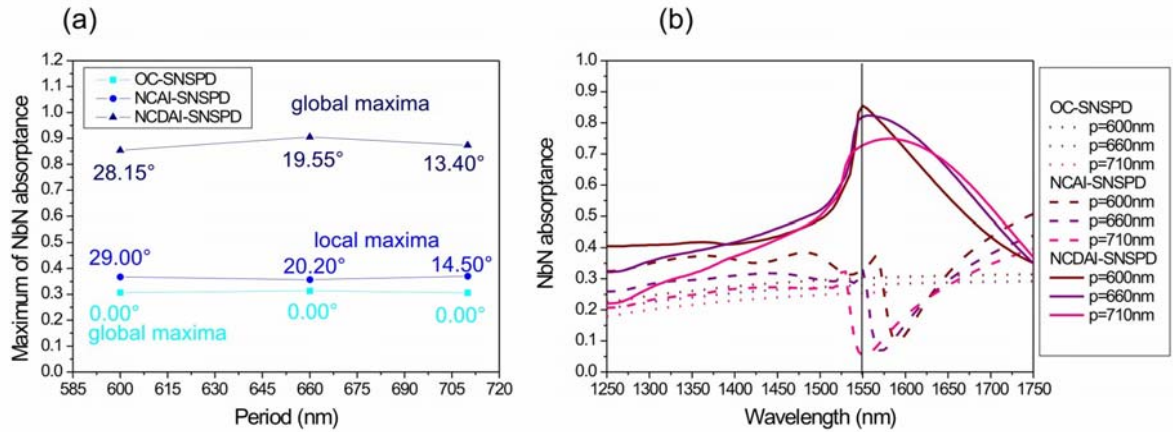

Figure 5. (a) The normalized absorptance in 3$p$-periodic SNSPD designs: (b) wavelength dependent absorptance at the orientations indicated in (a) in different designs.

Interestingly, the 660-nm-pitch NCAI-SNSPD exhibits the smallest normalized absorptance at the local maximum. In NCDAI-SNSPDs the global absorptance maxima (1550 nm, 1561 nm, 1585 nm) are in close proximity of the 1550 nm operation wavelength of SNSPDs. Even though the maximum is slightly detuned from 1550 nm in 660-nm-pitch design, the larger FWHM makes possible to attain larger normalized absorptance in practical applications.

**Near-field phenomena in short-periodic designs**

The near-field investigation has shown that there is an **E**-field antinode at the bottom of quarter-wavelength nano-cavities in each short-periodic OC-SNSPDs (Fig. 6a/a-c).

In short-periodic NCAI-SNSPDs the normalized **E**-field indicates intense antinodes at the bottom of each quarter-plasmon-wavelength nano-cavity at the orientations corresponding to the global maxima (Fig. 6b/a-c). As the global maximum appears at the tilting close to SPP excitation in 200-nm-pitch design, additional **E**-field enhancement is observable above the horizontal gold segment (Fig. 6b/a), and the time evolution of the **E**-field shows SPP propagation (Supplementary Video 1). There is no significant difference between the maximal values of the **E**-field, similarly the resistive heating as well as the reached absorptance maxima are almost the same in the three $p$ periodic NCAI-SNSPD design. This indicates that the efficiency of collective resonances on sub-wavelength periodic MIM nano-cavity arrays only slightly depends on the periodicity in the investigated $p$ periodicity interval.

According to this, there are no reflected waves below *p* periodic NCAI-SNSPD designs (Supplementary Video 1-3).

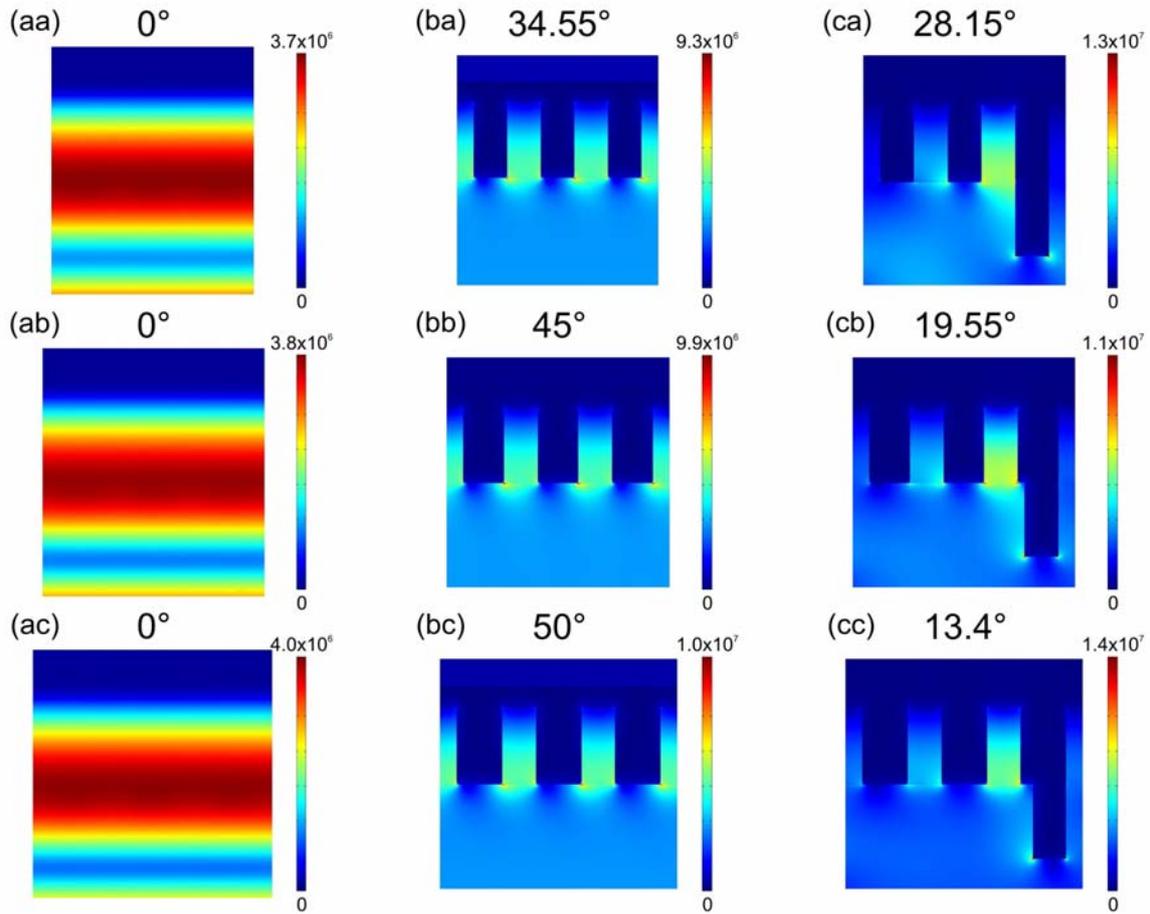

Figure 6. Near-field distribution in short-periodic SNSPD designs at the maxima: (a/a-c) OC-SNSPD, (b/a-c) NCAI-SNSPD, (c/a-c) NCDAI-SNSPD; first row p = 200 nm, second row p = 220 nm, third row p = 237 nm.

In device designs consisting of deflectors with 3*p* periodicity integrated into the *p* periodic intertwining bi-grating of NbN stripes and gold nano-cavity-array, the **E**-field is further enhanced significantly in nano-cavities in the closest proximity of the deflector (Fig. 6c/a-c). These deflectors not only perturb the **E**-field, they also absorb a part of the incoming light. As a consequence, smaller absorptance is attainable in short-periodic NCDAI-SNSPDs at orientations corresponding to the global maxima in their long-periodic pairs, then in NCAI-SNSPDs, i.e. without deflectors. Interestingly, surface waves with fronts perpendicular to the sapphire-interface are observable in all three *p* periodic NCDAI-SNSPDs at these orientations (Supplementary Video 4-6).

**Near-field phenomena in long-periodic designs**

Intense **E**-field antinodes are observable at the bottom of quarter-wavelength nano-cavities in each long-periodic OC-SNSPDs, similarly to their short-periodic pairs (Fig. 7a/a-ac).

In long-periodic NCAI-SNSPDs, in addition to the enhanced normalized **E**-field at the bottom of each quarter-plasmon wavelength MIM nano-cavity, **E**-field enhancement is observable also below the NbN segments due to the surface waves excitable at the orientations corresponding to the local maxima (Fig. 7b/a-c). In our previous work we have shown that the local maximum appears at the tilting corresponding to Brewster wave excitation in 600-nm-pitch design [15].

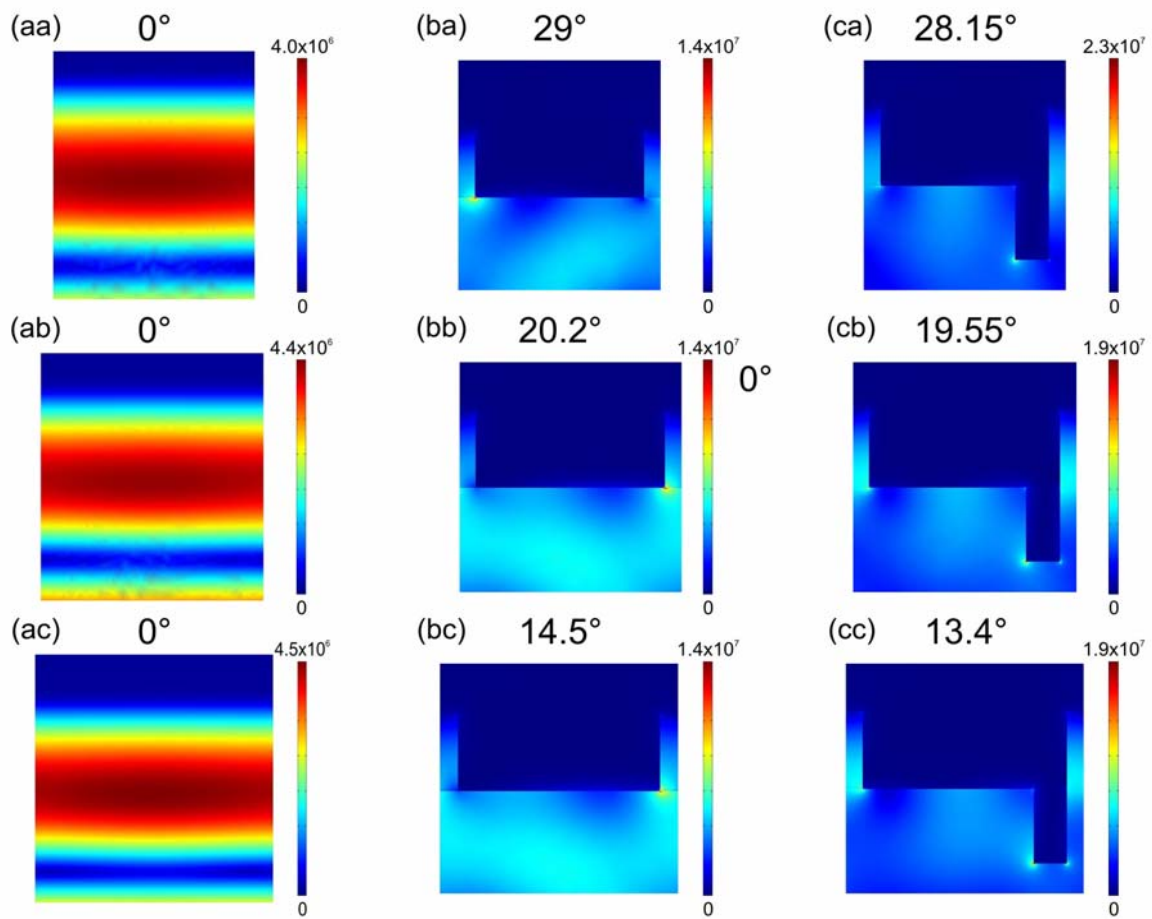

Figure 7. Near-field distribution in long-periodic SNSPD designs at the maxima: (a/a-c) OC-SNSPD, (b/a-c) NCAI-SNSPD, (c/a-c) NCDAI-SNSPD; first row $p =600$ nm, second row $p = 660$ nm, third row $p = 710$ nm.

Similar backward propagating Brewster surface waves having wavefronts perpendicular to the NCA-sapphire interface, and wavelength larger than the light wavelength in sapphire, are observable on the time evolution of the **E**-field in all three $3p$ periodic NCAI-SNSPDs (Fig. 7b/a-c, Supplementary Video 7-9). These near-field results prove that the propagating surface waves are capable of resulting in local absorptance enhancement, even though the efficiency of collective resonances on MIM nano-cavity arrays is smaller, and exhibits stronger dependence not only on the tilting, but also on the periodicity in the investigated $3p$ pitch interval. Interestingly, even though there is no significant difference between the maximal values of the **E**-field and similar surface waves are observable in the three systems, the normalized absorptance is smaller in 660-nm-pitch design, than in either of 600-nm and 710-nm-pitch designs.

In $3p$ periodic NCDAI-SNSPDs the time evolution of the **E**-field proved that the deflection of grating-coupled surface waves, with wavelength according to SPPs, by the longer vertical gold segment makes possible strong synchronous **E**-field enhancement also below the NbN segments in each MIM cavity (Fig. 7c/a-c, Supplementary Video 10-12). These results indicates that the **E**-field is confined above the NbN segments due to the resonant oscillation of nano-plasmonic modes in the MIM cavities, and below the NbN segments via propagating SPPs interacting with them in all three $3p$ periodic NCDAI-SNSPDs.

**Discussion**

In OC-SNSPDs the gold reflector covered quarter-photonic-wavelength dielectric cavity ensures that the **E**-field antinodes overlap with the NbN stripes (Fig. 6-7/a), which makes possible to attain absorptance higher than in bare NbN SNSPDs [11, 17]. The preferred $\gamma=0°$ azimuthal angle corresponding to P-orientation ensures the parallelism of the **E**-field oscillation direction to the NbN stripes, which is known to promote **E**-field penetration. Illumination with perpendicularly incident beams is the most advantageous in all OC-SNSPDs [Fig. 2-5/a], and the maximal absorptance decreases with the fill factor. The normalized absorptance monotonously increases at short ($p$) periodicities, while exhibits a local maximum at 660-nm-pitch, i.e. this design is proposed for practical applications.

In NCAI-SNSPDs the resonance in quarter-plasmonic-wavelength MIM cavities ensures that the high **E**-field intensity at the antinodes is coincident with the NbN segments lying at their entrance (Fig. 6-7/b) [18]. The **E**-field oscillation has to be perpendicular to the integrated gold segments, when plasmonic phenomena are at play, which explain that $\gamma=90°$ azimuthal angle corresponding to S-orientation is preferred. The coupled resonance on sub-wavelength periodic quarter-wavelength MIM plasmonic cavities results in polar-angle independent almost perfect absorptance in all *p*-pitch NCAI-SNSPDs (Fig. 2b) [19, 20], while the weak coupling in *3p* periodic NCAI-SNSPDs causes stronger dependence of the absorptance on tilting (Fig. 4b) [15, 18]. In all investigated *3p* periodic systems there are specific orientations, where the cavities are synchronously excited, making possible collective resonances. These orientations can be computed as:

$$\sin \varphi^{m,k} = \frac{m \frac{\lambda}{n_{sapphire}}}{kp}. \quad (1)$$

The observed local maxima appear at $m=1$, $k = 3, 4, 5$ cases in 600, 660, 710 nm-pitch-designs. As cavity resonances occur at 1550 nm illuminating light, plasmonic band-gaps and strong optical response modulation are observable in the intervals, where propagating surface modes are excited via grating-coupling. The wavelength of the modes that might be grating-coupled at the observed extrema can be calculated based on the momentum conservation: $k_{surface\ wave} = k_{photon} \pm k_{grating}$. The time evolution of the **E**-field shows backward propagating surface waves at the local maxima (Supplementary Video 1-3, 7-9), which originate from -1 order grating-coupling according to:

$$-\frac{2\pi}{\lambda_{surface\ wave}} = \frac{2\pi}{\lambda}\sin\varphi - \frac{2\pi}{3p}. \quad (2)$$

Based on this relationship the modes excited at the global minimum have wavelengths slightly smaller then the surface plasmon polaritons at unstructured sapphire-gold interface, while the modes at the local maximum have wavelengths larger then the photonic modes, revealing to their unique Brewster-mode nature.

At the local minima in all three 3$p$ periodic NCAI-systems, the simultaneous coupling and back-reflection at the SPP band-gap center prevents light-coupling into the cavities [8], while at the local maxima the synchronous excitation of backward propagating Brewster modes makes possible **E**-field enhancement (Fig. 6-7/b). The wavelength dependent study revealed that the absorptance enhancement is less effective in 660-nm-pitch design, where the local maxima coincide in polar angle and wavelength.

The NCDAI-SNSPDs are the designs, were the deflectors make possible to realize further **E**-field enhancement via propagating plasmonic modes, by back-deflecting them to the MIM nano-cavities, i.e. preventing their spurious back-reflection (Fig. 6-7/c, Supplementary Video 4-6 and 10-12). The presence of deflectors is not advantageous in short periodic NCDAI-SNSPDs in comparison to NCAI-SNSPDs, namely slightly smaller NbN absorptance is attainable caused by enhanced absorption on longer vertical segments, than the NbN absorptance due to collective resonances on sub-wavelength MIM cavity-arrays. The global maxima appear at smaller polar angles in 3$p$ periodic NCDAI-SNSPDs (Fig. 4c) than the local maxima in NCAI-SNSPDs (Fig. 4b), they are observable at the orientations where propagating plasmonic modes can be excited at gold-sapphire interface, according to Eq. (2). The observation of backward propagating surface modes proves that the presence of deflectors has real advantage, as they offer possibility to utilize the interaction of propagating and localized plasmonic modes in long-periodic designs (Fig. 7c and Supplementary Video 10-12). Illumination at the orientation corresponding to coupling into propagating plasmonic modes is proposed in these designs, and the compensation of losses caused by three-times smaller fill-factor might be realized.

Interestingly the 3$p$ periodicity interval is far from the wavelength of propagating plasmonic modes (~878 nm). The deviation between the periodicity and the wavelength, where extraordinary transmission is observable on plasmonic structures, was explained by the phase shift occurring upon reflection of plasmonic waves on wavelength-scaled periodic corrugations [21]. According to this, the observed phenomena are related to the Wood-Rayleigh phenomena [22].

The 660-nm-pitch equals to 0.75*$\lambda_{plasmon}$, i.e. this is the optimal periodicity, which is known to result in resonant transmission on plasmonic gratings, and simultaneously to minimize reflectance. Based on polar angle dependent study the peculiarity of 660-nm-pitch NCDAI-SNSPD design is that it makes possible resonant absorption due to resonant transmission and subsequent back-deflection of propagating plasmon on the integrated structure, as well as the resonant excitation of localized plasmons, and finally the most efficient interaction of propagating and localized modes due to proper synchronization. In addition to this, the wavelength dependent study revealed that the large absorptance maximum is accompanied with large bandwidth in 660-nm-pitch NCDAI-SNSPD design.

In conclusion, the optimal periodicity and illumination directions were determined for p-polarized light illumination of different SNSPDs designs. We have shown that the **E**-field concentration around the NbN segments at the bottom of gold reflector covered ~quarter-wavelength nano-optical cavities results in the highest absorptance at perpendicular incidence in P-orientation of OC-SNSPDs. Although the normalized absorptance decreases with the fill-factor in OC-SNSPDs, it indicates that 660-nm-pitch design is peculiar in 3*p* periodicity interval.

Our present study has confirmed that in NCAI-SNSPDs the collective resonances on MIM nano-cavity arrays in S-orientation strongly depend on the periodicity. Almost polar angle independent perfect absortpance is attainable in *p*-periodicity interval, while the 3*p* periodic integrated pattern acts as a plasmonic band-gap structure, where perpendicular incidence is the optimal, while contribution of surface waves is capable of resulting in local absorptance enhancements at specific orientations. Our present study has shown that these local maxima are due to Brewster waves. Interestingly the 660-nm-pitch, which is optimal for plasmonic band-gap engineering, results in relatively smaller normalized absorptance, caused by coincident local maxima in polar angle and wavelength.

We have proven that the re-radiation of plasmons, which causes the reflection of in-coupled light and inherits the absorption at the middle of the plasmonic gap, can be blocked by deflectors in NCDAI-SNSPDs.

The attainable absorptance strongly depends on the periodicity in these designs, namely the gold-deflector-related absorption loss overcomes the enhancement in *p*-periodic NCDAI-SNSPDs, while huge absorptance enhancement is reached at the middle of the band-gap due-to back-deflected plasmonic waves in 3*p* periodic patterns. Our present study has demonstrated that these deflectors force the surface plasmon polaritons to propagate backward, inhibit reflection and ensure synchronous high **E**-field intensity also below the absorbing NbN stripes. The discovered method is a general principle, which might be applied to eliminate radiation losses from nano-photonical systems based on plasmonic structures.

**Methods**

The FEM modeling was performed by the Radiofrequency module of Comsol Multiphysics software package (Comsol AB). The optimal illumination directions of each design were determined using a parametric sweep, namely by tuning the $\varphi$ polar and $\gamma$ azimuthal angles during p-polarized illumination of the SNSPD devices in conical mounting (Fig. 1d, e). First the optimal azimuthal orientation was selected by calculations performed over the $\gamma=[0-90°]$ and $\varphi=[0-85°]$ intervals, with $\Delta\gamma=\Delta\varphi=5°$ resolution. Then the optimal tilting was determined by varying the polar angle in $\varphi= [0°, 85°]$ region with 1° resolution. Finally higher (0.05°) resolution FEM calculations were performed in intervals surrounding the local/global maxima in order to inspect the accompanying near-field phenomena.

Wavelength dependent FEM computations were also performed at specific orientations of each designs corresponding to NbN absorptance maxima predicted by angle dependent studies. The wavelength dependent optical parameters were taken into account by integrating Cauchy formulae for dielectrics (HSQ ans sapphire), and loading tabulated data-sets for metals (NbN and gold) into the models. All calculations have a 5 nm resolution in the $\lambda=[1250-1750\ nm]$ region.

In the present numerical model computations NbN segments had 4 nm thickness and 100 nm width, and the ~2nm thick NbNOx cover-layer were taken into account (Fig. 1f). The geometry of the NbN pattern was also varied for each three designs, the three different periodicities inspected in details in the short *p*- and long 3*p*-periodicity regions were 200 nm, 220 nm, 237 nm in *p*, and 600 nm, 660 nm, 710 nm in 3*p* periodic patterns.

In OC-SNSPDs the ~quarter-wavelength nano-optical cavity filled with hydrogen-silsesquioxane (HSQ) was ~279 nm long and the single continuous cavity was closed by a 60 nm thick gold reflector (Fig. 1a).

In NCAI-SNSPDs the NbN stripes are aligned at the bottom of ~quarter-plasmon-wavelength, i.e. ~220 nm long nano-cavities, which are closed by 226 nm long vertical and 60 nm thick horizontal gold segments. The alternating HSQ and intersecting gold segments compose a plasmonic MIM cavity-array with *p* and 3*p* periodicity (Fig. 1b). The detailed description of the 200 nm and 600 nm NCAI-SNSPDs can be found in ref. [13-15], their polar-angle dependent absorptances are compared to the 220-nm and 237-nm-pitch designs, as well as to the 660-nm and 710-nm-pitch NCAI-SNSPs in present work.

In NCDAI-SNSPD the deflector segments with 3*p* periodicity integrated into both *p* and 3*p* periodic NbN patterns (Fig. 1c) were 220 nm long and 100 nm wide in all cases.

**Acknowledgement**

This work has been supported by the U.S. Dept. of Energy Frontier Research Centers program. The study was funded by the National Development Agency of Hungary with financial support form the Research and Technology Innovation Funds (OTKA CNK-78549), and OTKA K 75149. Mária Csete thanks the Balassi Institute for the Hungarian Eötvös post-doctoral fellowship.

**Author contributions**

Mária Csete contributed to the paper by proposing the novel concept of deflectors to maximize absorptance in SNSPDs integrated with plasmonic structures (NCDAI-SNSPDs), analyzed the results and wrote the manuscript. Áron Sipos, Anikó Szalai and Faraz Najafi prepared the models for numerical simulation, and performed the near-field analyses. Gábor Szabó contributed to the concept of the paper by suggesting the comparison of nano-photonical phenomena in different periodicity intervals. Karl K. Berggren contributed to the initial concept of the paper by suggesting the comparative study of OC- and NCAI-SNSPDs, and application of 3*p* periodicity interval for simultaneous electrical optimization.

**Additional information**

**Competing financial interests:** The authors declare no competing financial interests.